\documentclass[twocolumn,showpacs,amstex,amssymb,prl]{revtex4}
\usepackage{graphicx}

\begin{document}

\title{Nearsightedness of Electronic Matter}

\author{E. Prodan$^{1,2}$ and W. Kohn$^1$}

\address{
$^1$Dept. of Physics, University of California, Santa Barbara, CA
93106}

\address{
$^2$Dept. of Materials Science, University of Southern California,
Los Angeles, CA 90089}

\begin{abstract}
In an earlier paper, W. Kohn had qualitatively introduced the
concept of ``nearsightedness" of electrons in many-atom systems. It
can be viewed as underlying such important ideas as Pauling's
``chemical bond," ``transferability" and Yang's computational
principle of ``divide and conquer." It describes the fact that, for
fixed chemical potential, local electronic properties, like the
density $n(r)$, depend significantly on the effective external
potential only at nearby points. Changes of that potential, {\it no
matter how large}, beyond a distance $\textsf{R}$ have {\it limited}
effects on local electronic properties, which rapidly tend to zero
as function of $\textsf{R}$. In the present paper, the concept is
first sharpened for representative models of uncharged fermions
moving in external potentials, followed by a discussion of the
effects of electron-electron interactions and of perturbing external
charges.
\end{abstract}


\maketitle

It is a conventional qualitative wisdom among physicists and
chemists that, in the absence of long range ionic interactions, if
an atom A in a solid is exchanged for another atom B, the change of
the total energy of the system is largely determined by atoms A and
B and their near neighbors. Similarly for a molecule M adsorbed on a
surface. These are qualitative examples of ``nearsightedness."

Understanding the physics and chemistry of large molecules and
solids would have been practically impossible if not for the
principle of transferability \cite{Levy,Allen1}. It is generally
accepted that, in the absence of long range ionic interactions,
large molecules or materials systems can be studied and understood
one neighborhood at a time, without the necessity of studying the
entire system at once. The useful computational method of ``divide
and conquer" takes advantage of this fact \cite{Yang91}.

We can argue that Pauling's concept \cite{Pauling} of the chemical
bond has a well defined meaning because, to a good approximation,
its properties depend only on the relative positions of the bonded
atoms and their near neighbors \cite{Allen2}. Anything beyond them,
has little influence on the properties of the chemical bond.

These important concepts, based on decades of empirical and
computational work, point to a property of matter which we call
``nearsightedness of electronic matter (NEM)." It was first realized
and introduced by one of the authors (WK) in 1996 \cite{Kohn96}. In
retrospect, one can find precursors of NEM implicit in many other
contexts: For example, in the work of Thomas and Fermi in the 1920's
\cite{ThomasFermi}; in the proposal of the Local Density
Approximation in 1965 \cite{KohnSham} and in Yaniv and Kohn's paper
of 1979, "Locality Principle in Wave Mechanics" \cite{Yaniv}. It was
also noticed in Lang and Kohn's extensive work on surfaces
\cite{Lang} and as an element of the concept of ``edge electrons" in
Ref.~\cite{Mattsson}.

NEM deals with the following scenario: We consider an unperturbed
system of very many charged or uncharged electrons in equilibrium in
an external, static potential $v(r)$, with chemical potential $\mu$,
at $T=0^+$. We are interested in the effect (for fixed $\mu$) of a
perturbing potential (change of the external potential)
$w(r^\prime)$, of finite support (footprint) \cite{footnot}, on a
local property at a point $r_0$, like the density $n(r)$, when the
support of $w(r^\prime)$ is outside a sphere of radius $\textsf{R}$,
centered at $r_0$ (Fig.~1). The NEM principle states that, for a
given unperturbed system and a given $\textsf{R}$, the density
changes at $r_0$, $\Delta n(r_0)$, due to {\it all} admissible
$w(r^\prime)$, have a finite maximum magnitude, $\overline{\Delta
n}$, which, of course, depends on $r_0$, $\textsf{R}$, and on the
unperturbed system. From this definition, one can see that
$\overline{\Delta n}(r_0,\textsf{R})$ decays monotonically as a
function of $\textsf{R}$. In this paper we prove, for broad classes
of systems, that in fact,
\begin{equation}
    \lim_{\textsf{R}\rightarrow \infty} \overline{\Delta n}(r_0,R)=0,
\end{equation}
and expect this to be valid very generally. We shall show that, for
ordered gapless systems, the decay follows power-laws, for ordered
gapped systems the decay is exponential and for disordered, gapped
or ungapped systems, the decay is also exponential.

\begin{figure}
\begin{center}
  \includegraphics[width=8.6cm]{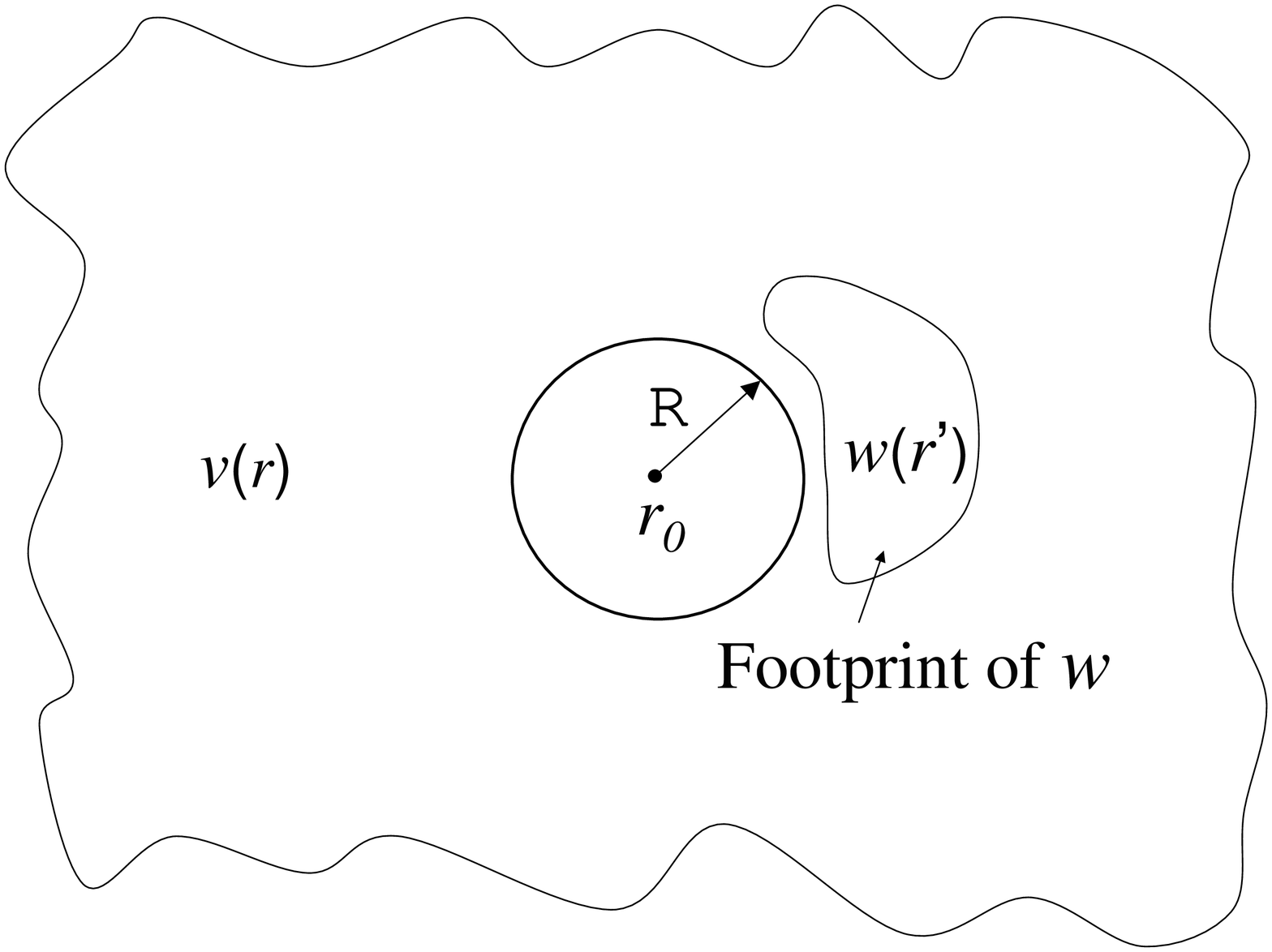}\\
  \caption{Schematic of Nearsightedness of Electronic Matter: $v(r)$
  is the unperturbed external potential, $w(r^\prime)$ is the
  perturbing potential outside a sphere of radius $\textsf{R}$,
  which is centered on the point of interest $r_0$ (see text for
  details).}
  \end{center}
\end{figure}

For a given $r_0$ and $\Delta n$, we can solve for $\textsf{R}$ from
$\overline{\Delta n}(r_0,\textsf{R})=\Delta n$ and hence define what
we call the nearsightedness range $\textsf{\textsf{R}}(r_0,\Delta
n)$. The significance of $\textsf{R}(r_0,\Delta n)$ is the
following: the density changes at $r_0$ due to any perturbation, of
arbitrary shape and amplitude, beyond $\textsf{R}(r_0,\Delta n)$,
cannot exceed $\Delta n$. We can say, anthropomorphically that, to
within an accuracy $\Delta n$, the particle density $n(r)$ cannot
``see" {\it any} perturbation $w(r')$ beyond the distance
$\textsf{R}(r_0,\Delta n)$; hence our word ``nearsightedness."

NEM frequently reminds of other well known and well understood
concepts but, in fact, is different. To avoid ``deadly sins," let us
present a list of what NEM is not:

1) NEM is not an aspect of linear or higher order non-linear
response to external perturbations (but does not exclude these).

2) NEM is not screening of charges, which renders long range Coulomb
potentials short range (NEM applies also to neutral fermions).

3) NEM does not apply to systems of few electrons or to
non-interacting bosons below their condensation temperature
(interacting bosons are beyond the scope of this paper).

4) NEM is not limited to electrons at $T=0^+$ but carries over to
finite $T$, including the classical (high $T$) limit.

5) NEM is not limited to macroscopically homogeneous systems. E.g.,
it applies to a point $r$ on an interface.

 In this article, we communicate the first quantitative results on
NEM for 1, 2 and 3D non-interacting, periodic electrons and
preliminary results for non-periodic and interacting electrons. We
shall see that, no matter how complicated or strong $w(r^\prime)$
is, far away from the perturbation, the change of electron density
has a universal form, which is completely determined by the
reflection coefficient, in 1D, or elements of the scattering matrix,
in 2 and 3D, evaluated at certain energies. NEM follows from the
fact that these coefficients cannot exceed a certain upper bound.
Based on these asymptotic estimates, we discuss the nearsightedness
range and present an application to linear scaling electronic
algorithms.

\textbf{Non-interacting fermions:} We emphasise again that NEM, as a
general principle, does not require interactions or screening. It is
due to the destructive interference of {\it density} (not wave) {\it
amplitudes} $n_{j}$ associated with the occupied single particle
eigenstates $\psi _{j}$.

{\it One dimension:} We begin with a model of 1D electrons in a
periodic potential $v(x)$ with inversion symmetry, at $T=0^+$. The
unperturbed Hamiltonian is [$\hbar =2m=1$]
\begin{equation}\label{h0}
H_{0}=-d^{2}/dx^{2}+v(x),\,\,\,v(x+b)=v(x).
\end{equation}
We first restrict the perturbing potential $w(x)$ to vanish for
$x>0$. The density change is given by
\begin{equation}\label{prima}
    \Delta n(x)=\frac{1}{\pi i}\int_{\cal C} [G_E(x,x)-G_E^0(x,x)]dE,
\end{equation}
where $G_E^0$ and $G_E$ are the unperturbed and perturbed Green's
functions, respectively, and ${\cal C}$ is a contour surrounding the
eigenvalues below $\mu$. The integral in Eq.~(\ref{prima}) can be
mapped into the complex $k$-plane,
\begin{equation}\label{fundamental}
    \Delta n(x)=2\int_{\tilde{\cal C}} R(k)\psi_k(x)^2dk,\ \ x>0,
\end{equation}
where $\tilde{{\cal C}}$ corresponds to ${\cal C}$, $R(k)$ is the
reflection coefficient from right to left and $\psi_k(x)$ are the
normalized, unperturbed Bloch functions \cite{Prodan}.

\begin{figure}
\begin{center}
  \includegraphics[width=8.0cm]{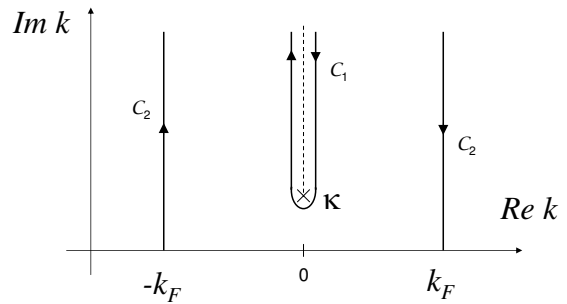}\\
  \caption{The contours ${\cal C}_1$ and ${\cal C}_2$ on
  the Riemann sheet of the last occupied band (assumed odd here).
  ${\cal C}_1$ surrounds the branch point $\kappa$ and the corresponding branch cut
  (dashed line). ${\cal C}_2$ intersects the real axis at $k_F$ and reappears at $- k_F$.
  }
  \end{center}
\end{figure}

{\it a) Asymptotics:} For insulators, we can and shall restrict
ourselves to $\text{Im} k \geq 0$ and to the first Brillouin zone.
We denote by $\kappa$ the branch point that connects the highest
occupied and the lowest unoccupied bands \cite{Kohn59}. For the case
when $w(x)$ generates no bound states in the insulating gap, we can
choose $\tilde{\cal C}$ in Eq.~(\ref{fundamental}) to be the contour
${\cal C}_1$ in Fig.~2. Since $\psi_k(x)=u_k(x)e^{ikx}$, with $u_k(x)$ periodic of $x$, $\psi_k(x)$ decays exponentially with
$x$ ($\sim e^{-\text{Im}kx}$) for $\text{Im}k>0$ and the asymptotic
behavior of $\Delta n(x)$ comes from the points of ${\cal C}_1$ in
the immediate vicinity of $\kappa$. Using the behavior of the Bloch
functions near the branch point \cite{Kohn59}, we find
\begin{equation}
    \Delta n(x) \rightarrow 2R(\kappa) \left( \frac{2\pi
    }{x}\right) ^{1/2}s_{\kappa}(x)^{2}e^{-2qx},  \label{asymptotic1}
\end{equation}
where $q\equiv \text{Im}\kappa$ and $s_{\kappa}(x)\equiv [|k-\kappa
|^{1/4}u_{k}(x)]_{k\rightarrow \kappa }$ is a real, periodic or
antiperiodic function, $s_\kappa(x+b)=\pm s_\kappa(x)$, depending on
the band index. $w(x)$ enters in this asymptotic form only through
$R(\kappa)$.

The sign of this $\Delta n(x)$ is independent of $x$ and is given by
the sign of $R(\kappa)$. The exponential decay constant $q$ in
Eq.~(\ref{asymptotic1}) equals that of the Wannier function of the
highest occupied band \cite{OnffroyKohn73,RehrKohn74,Pei77}, or of
the density matrix \cite{Vanderbilt01}. In the limit of a
sufficiently small insulating gap $G$ (in units of the width $W$ of
the highest occupied band), $q=\frac{1}{2}\sqrt{m^* G},$ where $m^*$
is the effective mass at the top of the last occupied band
\cite{footnote0}.

One can show that $|R(\kappa)|_{\max}=1$, with the maximum taken over all functions
$w(x)$ which generate no bound states in the insulating gap
\cite{Prodan}. Thus, $\Delta n(x)$ cannot exceed an upper bound,
independent of $w$ (If there are bound states, the asymptotic
behavior of $\Delta n(x)$, Eq.~(\ref{asymptotic1}), may change, but
NEM remains \cite{Prodan}).

For metals, we integrate Eq.~(\ref{fundamental}) along the contour
${\cal C}_2$ in Fig.~2. For large $x$,
\begin{equation}\label{asymptotic2}
    \Delta n(x) \rightarrow \frac{2}{x}\text{Im}[R(k_F)\psi
    _{k_F}(x)^{2}],
\end{equation}
the slowly decaying Friedel oscillations \cite{Friedel52}. Again,
NEM follows from the fact that $|R(k_F)|$ cannot exceed 1, for any
$w(x)$.

For perturbed Kronig-Penney models \cite{Kronig}, we found that the
asymptotic expressions Eqs.~(\ref{asymptotic1}) and
(\ref{asymptotic2}) set in after one or two lattice parameters.

{\it b) Nearsightedness range} \textsf{R}{\it :} For a given $\Delta
n$, the nearsightedness range $\textsf{R}(r_0,\Delta n)$ at $r_0$
was introduced as the smallest distance such that any scalar
perturbation $w(r^\prime)$, lying entirely outside this range,
produces a density change at $r_0$, $\Delta n(r_0)$, smaller than
$\Delta n$. Fully characterized, \textsf{R} is a function of $r_0$,
chemical potential $\mu$ and $\Delta n$, and a functional of $v(r)$
\begin{equation}
    \textsf{R}\equiv \textsf{R}\left(r_0,[v(r')],\mu,\Delta n \right).
\end{equation}
From this definition, it follows that, at every fixed $r_0$,
$\partial \textsf{R}(r_0,\Delta n)/\partial \Delta n\leq0.$

To calculate $\textsf{R}$ at a point $x_0$, we need to
simultaneously consider perturbing potentials $w_{L,R}$ to the left
and right of $x_0$. In this case, the density change at $x_0$ is
given by the individual contributions $\Delta n_{L,R}(x_0)$ of
$w_{L,R}$, Eqs.~(\ref{asymptotic1}) and (\ref{asymptotic2}), plus
multiple reflections corrections. In the limit when the distance
from $x_0$ to $w_{L,R}$ is large, these corrections were found
exponentially small for insulators and comparable to $\Delta
n_{L}(x_0)+\Delta n_{R}(x_0)$ for metals.

For insulators, the asymptotic behavior of $\textsf{R}$ in the limit
$\Delta n \rightarrow 0$, as derived from the upper bound of
Eq.~(\ref{asymptotic1}), from the above remark and from a
cell-averaging of $\Delta n(x)$, is
\begin{equation}\label{near1}
    \textsf{R}(\Delta n)\rightarrow\frac{1}{2q}\ln \frac{\tilde{n}}{\Delta n},
\end{equation}
where
\begin{equation}\label{tilden}
    \tilde{n}=\frac{8\sqrt{2 \pi q}}{b}\int_{0}^{b}s_{\kappa}(x)^2 \ dx.
\end{equation}
In the small gap and tight binding limits, $\tilde{n}$ is completely
determined by the exponential decay constant $q$,
$\tilde{n}\rightarrow 4q\sqrt{2/\pi}$ and by $\tilde{n}\rightarrow
4\sqrt{q/\pi b}$, respectively.

For metals, the upper bound of Eq.~(\ref{asymptotic2}), with
inclusion of multiple reflections and cell-averaging, leads to the
following asymptotic behavior
\begin{equation}\label{near2}
    \textsf{R}(\Delta n)\rightarrow1/\Delta n.
    \label{neasightedness2}
\end{equation}

{\it c) Disorder:} NEM is {\it not} limited to periodic $v(x)$. As a
first orientation, we discuss the effect of adding a small random
potential
\begin{equation}
    v_{r}(x)=\lambda\sum_{n}\delta(x-x_n),
\end{equation}
to the periodic potential $v(x)$ in Eq.~(\ref{h0}). We have
calculated the averaged density-density correlation function, to the
lowest order in $\lambda$, assuming a random distribution of
impurities with an average density $n_i$ (cf. \cite{Edwards}). Our
conclusions are as follows.

For insulators, the random potential changes the exponential decay
constant $q$ by
\begin{equation}
    \Delta q=-n_i \lambda^2\frac{3(4\pi)^2}{2b\gamma ^2}
    \int_0^b [s_{-\kappa}(x)s_{\kappa}(x)]^2 \ dx,
\end{equation}
where $\gamma=[(E_k-E_{\kappa})/(k-\kappa)^{1/2}]_{k\rightarrow
\kappa}$. Two thirds of $\Delta q$ are due to the narrowing of the
gap $G$,
\begin{equation}
    \Delta G=-n_i
    \lambda^2\frac{2(4\pi)^2}{bG}\int_{0}^{b}[s_{-\kappa}(x)s_{\kappa}(x)]^2 \ dx,
\end{equation}
and one third is due to the fluctuations.

For metals, the random potential introduces an exponential decay in
the averaged density response, with a decay constant
\begin{equation}
    q=n_i \lambda^2 g(\epsilon_F)^2 \frac{12 \pi ^4}{b} \int_0^b
    |\psi_{k_F}(x)|^4 dx,
\end{equation}
where $g(\varepsilon_F)$ is the density of states at the Fermi
energy.

Thus, a small random potential increases $\textsf{R}$ for insulators
and decreases it for metals. We conjecture that this remains true
for disorder, more generally.

{\it Higher dimensions:} We first consider, as an example, 2D
fermions in a periodic potential with square symmetry. We present
only the cases when the first band, assumed simple and isolated
\cite{footnote1}, is partially or completely filled. We first
restrict the perturbing potential $w(x,y)$ to vanish for $y>0$ and
the periodicity along the $x$ direction to be preserved (so that
$k_x$ remains a good quantum number). The density change for
$y\rightarrow \infty$ is given by \cite{Prodanp}
\begin{equation}\label{compact}
    \Delta n(\vec{r})
    \rightarrow 2\int\limits_{\epsilon_{\vec{k}}<\epsilon_F}S(\vec{k},\vec{k}^\prime)
    \psi_{\vec{k}}(\vec{r})\psi_{\vec{k}^\prime}^\ast(\vec{r}) d\vec{k},
\end{equation}
with $\vec{k}^\prime=(k_x,-k_y)$ and $S(\vec{k},\vec{k}^\prime)$ the
scattering matrix element between $\vec{k}$ and $\vec{k}^\prime$.

{\it a) Asymptotics:} For fixed $k_x$, the analytic structure
relative to $k_y$ of the band energy $E_{\vec{k}}$ and Bloch
function $\psi_{\vec{k}}$ of the first band is completely analogous
to that in 1D \cite{Kohn59}: $E_{\vec{k}}$ and $\psi_{\vec{k}}$ have
branch points of order 1 and 3, respectively, at $\kappa_y=\pm
\pi+iq(k_x)$, connecting the first band with a higher band (we
restrict ourselves to $\text{Im}k_y\geq0$ and to the first Brillouin
zone). Their behavior near these points is the same as in 1D, namely
$E_{\vec{k}}$ behaves as a square root and $\psi_{\vec{k}}$ diverges
as $(k_y-\kappa_y)^{-1/4}$ \cite{Prodanp}.

If the band is completely filled and $w(\vec{r})$ does not generate
bound states in the insulating gap, for a given $k_x$, the integral
over $k_y$ in Eq.~(\ref{compact}) can be taken over a contour
surrounding the branch point (as in Fig.~2) and its asymptotic
behavior can be determined as in the 1D case. Thus, the integral
over $k_y$ decays exponentially, as function of $y$, with a rate
$2q(k_x)$. There will be two values, $\pm k_x^0$, of $k_x$ where
$q(k_x)$ reaches its lowest value $q_0$. The asymptotic behavior of
$\Delta n(\vec{r})$, for $y\rightarrow \infty$, comes from the
immediate vicinity of these points. Defining $\beta\equiv(\partial^2
q/\partial^2 k_x)_{k_x=k_x^0}$ and
$\vec{\kappa}_0\equiv(k_x^0,\pi+iq_0)$,
$\vec{\kappa}^\prime_0\equiv(k_x^0,-\pi-iq_0)$, gives
\begin{equation}\label{asymptotic3}
    \Delta n(\vec{r})\rightarrow
    \frac{4\pi}{y}\sqrt{\frac{2}{\beta}}\text{Re}[S(\vec{\kappa}_0,\vec{\kappa}^\prime_0)
    \phi_{\vec{\kappa}_0}(\vec{r})\phi_{\vec{\kappa}_0^\prime}^\ast(\vec{r})]e^{-2q_0y},
\end{equation}
where $\phi_{\vec{\kappa}_0}(\vec{r})\equiv
[|k_y-\kappa_y|^{1/4}e^{-ik_yy}\psi_{\vec{k}}(\vec{r})]_{\vec{k}\rightarrow
\vec{\kappa}_0}$ is a quasi-periodic function in $x$ {\it and} $y$.
Again, $w(\vec{r})$ enters in this asymptotic form only through
$S(\vec{\kappa}_0,\vec{\kappa}^\prime_0)$.

The exponential decay in Eq.~(\ref{asymptotic3}) is twice as fast as
the exponential decay, in the $y$ direction, of the density matrix
or of the Wannier function of the first band.
$|S(\vec{\kappa}_0,\vec{\kappa}^\prime_0)|=1$ for a hard wall and,
in general, we expect it to be of order 1.

If the band is partially filled, the asymptotic behavior of $\Delta
n(\vec{r})$ is determined by the two points on the Fermi surface,
denoted by $\vec{k}_0\equiv(k_x^0,k_y^0)$ and
$\vec{k}_0^\prime\equiv(k_x^0,-k_y^0)$, where the tangent to the
Fermi surface is along the $k_x$-direction:
\begin{equation}\label{asymptotic4}
    \Delta n(\vec{r})\rightarrow 2\text{Im}\sqrt{\frac{-i\pi}{\eta
    y^3}}S(\vec{k}_0,\vec{k}_0^\prime)
    \psi_{\vec{k}_0}(\vec{r})\psi_{\vec{k}_0^\prime}^\ast(\vec{r}),
\end{equation}
with $\eta$ the curvature of the Fermi surface at these points. From
the unitarity of the scattering matrix, one can immediately find
that $|S(\vec{k}_0,\vec{k}_0^\prime)|_{\max}=1$, with the maximum
taken over all $w(\vec{r})$. Thus, the asymptotic density change
cannot exceed an upper bound, no matter how large the perturbing
potential is.

The 3D case is analogous.

{\it b) Nearsightedness range} \textsf{R}{\it :} For metals, the
simplest model is jellium enclosed by a spherical hard wall, for
which one easily finds
\begin{equation}
    \textsf{R}(\Delta n)\rightarrow
\left \{
    \begin{array}{c}
    k_F/2 \Delta n=2.2 r_s \bar{n}/\Delta n \ \ (2D) \\
    k_F^2/2\pi\Delta n =2.5 r_s \bar{n}/\Delta n\ \ (3D),\\
    \end{array}
    \right.
\end{equation}
where $\bar{n}$ is the density of the uniform gas and $r_s$ the
Wigner-Seitz radius. For metals in periodic potentials, the $\Delta
n$ dependence remains unchanged, but $k_F$ is replaced by a
$k_{eff}$ depending on the band structure and the filling.

For an insulator with square symmetry, we calculated $\Delta
n(\vec{r_0})$ due to enclosing the point $r_0$ in four hard walls
along the symmetry axes, at a distance $D$ from $r_0$. The density
change near $r_0$ is given by the sum of the changes due to each
individual, infinitely extended wall, Eq.~(\ref{asymptotic3}), plus
multiple reflection corrections, which were found to be
exponentially negligible in the limit $D\rightarrow \infty$.
Similarly for a 3D insulator with cubic symmetry. From
Eq.~(\ref{asymptotic3}), its 3D analog \cite{Prodanp} and the
previous remark, we find that the cell-averaged density change at
$r$ becomes less than a given $\Delta n$ for $D\geq
\textsf{D}(\Delta n)$,
\begin{equation}\label{near3}
\textsf{D}(\Delta
n)\rightarrow\frac{1}{2q_0}\ln\frac{\tilde{n}}{\Delta n},
\end{equation}
where $\tilde{n}$ can be easily calculated from the band structure.

Finding an analytic expression of the nearsightedness range
$\textsf{R}$ for a general 2 or 3D insulator is clearly a next to
impossible task. However, on the basis of the above calculations,
the proof of exponential localization of 2 and 3D Wannier functions
\cite{Cloizeaux,Heine} and of 1D generalized Wannier functions
\cite{OnffroyKohn73, RehrKohn74}, as well as our 1D result, Eqs.
(\ref{asymptotic1}) and (\ref{near1}), we expect results of the
following form for 2 and 3D insulators:
\begin{equation}\label{neargeneral}
    \textsf{R}(\Delta n)\rightarrow\frac{1}{2q_{eff}}\ln\frac{\tilde{n}}{\Delta
    n},
\end{equation}
where $q_{eff}$ is an exponential decay constant of the density
matrix and $\tilde{n}\propto q_{eff}^d$ ($d=2,3$).

{\it c) Disorder:} We expect the effects of disorder in 2 and 3D to
be qualitatively similar to those in 1D.

 \textbf{Linear scaling:} The CPU time for electronic
structure calculations of a system consisting of many ($N_{a}$)
atoms grows very rapidly with $N_{a}$, if the calculations are
performed for the entire system at once. It has been pointed out
\cite{Yang91,Wu02} that the dependence on $N_{a}$ can be made linear
for large $N_{a}$, by dividing the system into $N_{s}$ suitable
sub-systems, where $N_s \propto N_a$. In Ref.~\cite{Kohn96}, NEM was
identified as the physical basis of linear scaling. Here, we
quantify this idea.

\begin{figure}
\begin{center}
  \includegraphics[width=7.5cm]{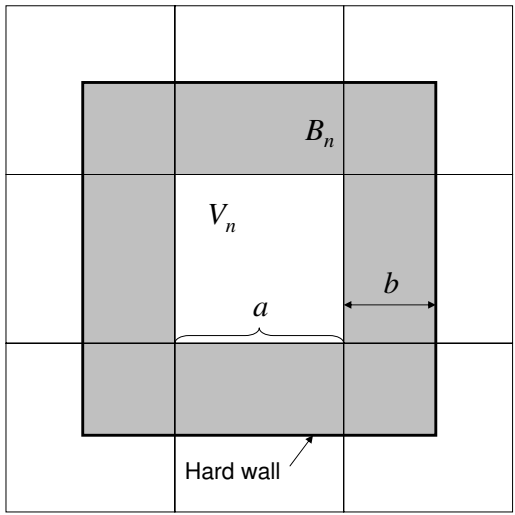}\\
  \caption{The system is divided into smaller volumes
  $V_n$ (9 in this example), with buffer zones $B_n$ (gray color).
  }
  \end{center}
\end{figure}

The procedure is illustrated in Fig.~3 (a detailed discussion
(including self-consistency) can be found in Ref. \cite{Wu02}). The
system of volume $V$ is divided into segments $V_n$, with
overlapping buffer zones $B_n$. The density $n(r)$ and contribution
to the total energy of each $V_n$ are obtained from calculations
including its buffer zone. For a required accuracy $\Delta n$, the
thickness $b$ of $B_n$ is chosen so that, when $r$ is on the
boundary of $V_n$, $\Delta n(r)\leq \Delta n$, where $\Delta n(r)$
is the error due to the hard walls around $B_n$. An upper bound for
$b$ is given by the maximum of $\textsf{R}(r,\Delta n)$ on the
boundary of $V_n$. Finally, the size of each $V_n$ is chosen to
minimize the total CPU time.

We exemplify this for ``cubic" periodic systems in 1, 2 and 3D,
where the $V_n$ are all identical ``cubes" of edge size $a$. The
largest density change occurs at the ``corners" of $V_n$ and comes
primarily from the nearest walls. For metals,
Eqs.~(\ref{asymptotic2}), (\ref{asymptotic4}) and their 3D analog,
plus the inclusion of the multiple reflections, lead to
$b\rightarrow \chi r_s (\bar{n}/\Delta n)^{2/(d+1)}$, with
$\chi=0.31$, 0.91 and 1.0 for 1, 2 and 3D, respectively. For a 5\%
accuracy, $b=6.2$, $6.7$ and $4.5r_s$, respectively. Similarly, for
insulators, $b\rightarrow (2q_0)^{-1}\ln[\tilde{n}/2\Delta n]$, with
$\tilde{n}$ defined in Eqs.~(\ref{tilden}) and (\ref{near3}).

The total CPU time is given by $t=N_s \tau$, where $N_s$ ($=V/a^d$)
is the number of segments, and $\tau$ ($\propto(2b+a)^{\nu d}$) is
the CPU time for the electronic structure calculation of one segment
plus its buffer zone [$\nu=2-3$ for DFT \cite{Wu02}, and higher for
other methods]. Minimizing the total CPU time with respect to $a$,
we obtain the optimal size, $a=2b/(\nu-1)$. With this optimization,
and from our estimates of $b$, we obtain the following dependence of
the total CPU time on the desired accuracy and the total number of
atoms:
\begin{equation}
    t\propto N_a \times \left\{
    \begin{array}{l}
    (\Delta n)^{2(1-\nu)d/(d+1) \ } \ \   \text{(ungapped)} \\
    \left(\ln \tilde{n}/2\Delta n \right )^{(\nu -1)d} \ \
    \text{(gapped)}
    \end{array}
    \right.
\end{equation}
For metals, $b$ can be greatly reduced by averaging the wall-induced
Friedel oscillations over two or more values of $b$.

\textbf{Interacting fermions:} In considering the response of
charged fermions to distant disturbances it is necessary to
distinguish between two cases:

a) Distant perturbing {\it potentials}, $w(r^\prime)$, with
$|r_0-r^\prime|\geq \textsf{R}$. The simplest description of
many-body interaction effects is the random phase approximation
which, in all dimensions, leads to a decrease of $\textsf{R}$ in
typical metals but an increase of $\textsf{R}$ in typical insulators
due to a reduction in the gap \cite{Prodan}.

b) Distant perturbing {\it charge densities} $\rho (r^\prime)$. In
analogy with $\textsf{R}(r_0,\Delta n)$, we define a
charge-nearsightedness range, $\textsf{R}_c(r_0,\Delta n)$ as the
smallest distance such that any charge perturbation $\rho(r^\prime)$
lying entirely outside this range produces a density change at
$r_0$, $\Delta n(r_0)$, smaller than $\Delta n$.

As is well known, the long range Coulomb potential, due to
perturbing electric charges, is screened out by metallic electrons.
Preliminary model calculations for metallic electrons, in the
Thomas-Fermi approximation, indicate that they are
charge-nearsighted, i.e. have a finite $\textsf{R}_c$. However,
charged {\it insulating} fermions are "classically farsighted," in
the sense that, at sufficiently large distances, the fermions ``see"
the classical long range total potential $\int
\rho_t(r^\prime)/|r_0-r^\prime|dr^\prime$, were $\rho_t$ is the
total perturbing charge density, including depolarization. Thus, for
example, in metals, replacing a neutral atom or ion by another atom
or ion has always short range electronic consequences, while, in an
insulator, ions lead to classical long range electronic effects.

This work was supported by Grants No. NSF-DMR03-13980,
NSF-DMR04-27188 and DOE-DE-FG02-04ER46130. One of us (W.K.)
gratefully acknowledges the frequent hospitality of the Institute
for Theoretical Physics of the ETH, Zurich, and stimulating
discussions with Prof. J. Friedel.

\end{document}